\documentclass{aa}
\usepackage{graphicx,natbib,amsmath,amssymb,fnpara}
\usepackage{txfonts}
\bibpunct[, ]{(}{)}{,}{a}{}{,}

\begin{document}

   \title{The X-ray afterglow of GRB\,020322}
%   \titlerunning{}

   \author{D.~Watson\inst{1}
          \and
          J.~N.~Reeves\inst{1}
	  \and
          J.~P.~Osborne\inst{1}
          \and
%          K.~A.~Pounds\inst{1}
%          \and
          J.~A.~Tedds\inst{1}
          \and
          P.~T.~O'Brien\inst{1}
          \and
          L.~Tomas\inst{2}%\fnmsep\thanks{Just to show the usage of the elements}
          \and
          M.~Ehle\inst{2}%\fnmsep\thanks{Just to show the usage of the elements}
          }

   \offprints{D.~Watson, email: \texttt{wat@star.le.ac.uk}}

   \institute{X-ray Astronomy Group, Dept. of Physics and Astronomy, University of Leicester, Leicester LE1 7RH, UK
         \and
             XMM-Newton Science Operations Centre, European Space Agency, P.~O.~Box 50727, 28080 Madrid, Spain
             }

   \date{Received ; accepted }

   \abstract{The spectrum of the afterglow of GRB\,020322 is the highest-quality
             X-ray spectrum of a GRB afterglow available to date.  It was
             detected by \emph{XMM-Newton} in an observation starting
             fifteen hours after the GRB with a mean 0.2--10.0\,keV observed
             flux of $3.5\pm0.2\times10^{-13}$\,erg\,cm$^{-2}$\,s$^{-1}$,
             making it the brightest X-ray afterglow observed so far with
             \emph{XMM-Newton}.  The source faded; its lightcurve
             was well fit by a power-law with a decay index of
             $1.26\pm0.23$.  The spectrum is adequately fit with a power-law
             absorbed with neutral or ionised gas significantly in excess of
             the foreground Galactic column, at redshift $1.8_{-1.1}^{+1.0}$
             or with low metal abundances.  No spectral line or edge
             features are detected at high significance, in particular, a
             thermal emission model fits the data poorly, the upper limit on
             its contribution to the spectrum is
             $3.7\times10^{-14}$\,erg\,cm$^{-2}$\,s$^{-1}$, or
             $\sim10\%$ of the total flux.  No spectral variability is
             observed.
   }

   \maketitle

%
%--------INTRODUCTION---------
%
\section{Introduction\label{introduction}}
%% Why X-ray afterglows are so much
%%   more useful than optical, i.e.
%%   line emission and X-ray 
%%   penetrating power.

Gamma-ray burst (GRB) afterglows are most often detected at X-ray energies
\citep{2001grba.conf...97P}, the majority of bursts producing no detectable
optical afterglow emission \citep{2001A&A...369..373F}.  Most redshift
estimates, however, are made from the absorption spectrum of the optical
transient or the apparent host galaxy.  Recently however, emission lines
have been detected in X-ray afterglows, allowing estimates to be made of the
cosmological redshifts and the outflow velocities of at least the
line-emitting component of the afterglow material
\citep{2000Sci...290..955P,2002Natur.416..512R,2002A&A...393L...1W}. 
In particular, the fact that metal-enriched thermal emission models fit the
highest-signal afterglow spectra better than power-law models
\citep{2002Natur.416..512R,2002A&A...393L...1W} has led to the strengthening
of the case for the association of long duration GRBs with the recent
collapse of a massive star
\citep{1993ApJ...405..273W,1998ApJ...507L..45V,2001ApJ...556..471L} and to
the assertion that thermal emission may be common in GRB afterglows
\citep{2002A&A...393L...1W}.  \emph{XMM-Newton} \citep{2001A&A...365L...1J}
has managed relatively rapid responses to GRB
alerts (on the order of half a day) and, with its large effective area, has
captured the high-quality spectra that have allowed these detections to be
made.

Previous detections of emission lines in GRB afterglows with
\emph{BeppoSAX}, \emph{ASCA} and \emph{Chandra} have concentrated on
emission from highly-ionised iron
\citep{1998A&A...331L..41P,2000Sci...290..955P,2000ApJ...545L..39A,2001ApJ...557L..27Y};
however the recent observations with \emph{XMM-Newton} have shown emission
features at lower energies (due in part to its very good sensitivity in the
soft X-ray band).

In Sect.~\ref{observations} we report on the observations of the afterglow
of GRB\,020322, explaining the data reduction procedure in
Sect.~\ref{reduction} and presenting the spectrum and lightcurve in
Sect.~\ref{results}.  In Sect.~\ref{discussion} the implications of these
results are discussed.  Conclusions are in Sect.~\ref{conclusions}.  Unless
otherwise stated, all errors quoted are 90\% confidence limits for one
parameter of interest, upper and lower limits are 99.7\% ($3\,\sigma$)
confidence limits and coordinates are equatorial, J2000.

%
%--------OBSERVATIONS---------
%
\section{Observations of GRB\,020322\label{observations}}

\begin{figure}
 \includegraphics[width=\columnwidth,clip=]{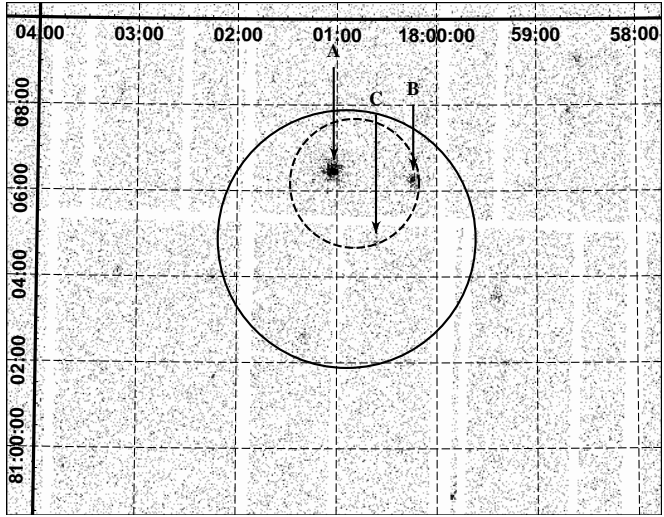}
 \caption{EPIC-pn 0.2--10\,keV-band image of GRB\,020322.  The \emph{BeppoSAX}
          WFC (solid circle -- 3\arcmin\ radius) and MECs (dashed circle --
          1.5\arcmin\ radius) error-circles are shown.  Three sources are
          detected within the MECS error-circle, labelled `A', `B' and `C';
          `A' is the only fading source in the EPIC data and is taken to be
          the X-ray afterglow of the burst.  Coordinate axes plotted are
          equatorial, R.A. and Dec.  (J2000).}
 \label{image}
\end{figure}

GRB\,020322 was detected by \emph{BeppoSAX} with the wide field camera (WFC)
on 22 March 2002, 03:51:30\,UT \citep{2002GCN..1290....1P} and followed 7.5
hours later with a detection in the MECS instrument
\citep{2002GCN..1291....1G}.  The position, (MECS: RA = 18h\,00m\,49.4s, Dec.\
= +81\degr\,06\arcmin\,10.8\arcsec), was well-determined, with an
error-radius of 3\arcmin\ in the WFC and 1.5\arcmin\ in the MECS (see
Fig.~\ref{image}).

\emph{XMM-Newton} began observing this position at 18:46\,UT with EPIC-MOS
and 19:17\,UT with EPIC-pn, approximately fifteen hours after the burst,
\citet{2002GCN..1293....1E} reporting detection of a bright source with a
positional accuracy of $\sim6$\arcsec.\ A fading optical source was detected
within this initial \emph{XMM-Newton} error-circle
\citep{2002GCN..1294....1B,2002GCN..1298....1G,2002GCN..1299....1M} at
RA = 18h01m\,02.98s, Dec. = +81\degr\,06\arcmin\,28\arcsec.17
\citep{2002GCN..1296....1B} with R-band magnitudes of $23.26\pm0.32$ and
$23.80\pm0.30$ at 10:39\,UT and 23:46\,UT respectively, suggesting an
early-time power-law decay slope of $\sim0.5$. The field was imaged with the
STIS instrument on \emph{HST} on 8 April and 5 May 2002, no optical
transient was detected  \citep[implying a late-time power-law decay slope steeper
than 2.0,][]{2002GCN..1536....1B}, however a galaxy of magnitude
$\sim27$ was detected at the position of the optical transient, probably the host.
No redshift has so far been determined for the GRB or the host galaxy.

\section{\emph{XMM-Newton}-EPIC data reduction and analysis\label{reduction}}

Effective exposure durations were 24\,ks and 28\,ks for the EPIC-pn
\citep{2001A&A...365L..18S} and MOS \citep{2001A&A...365L..27T} cameras
respectively, each in full frame mode using the thin filters. The data were
processed and reduced with the SAS, version 5.3, datasets from both EPIC-MOS
cameras were co-added.  MOS and pn spectra were fit individually, yielded
consistent results and were therefore fit simultaneously; the lightcurves
were compared for each instrument and were, again, consistent and were
therefore co-added and fit as a single dataset. Source extraction regions
were 40\arcsec\ in radius and off-source background extraction regions of
80\arcsec\ radius were chosen.  Both single and double pattern (as well as
triple and quadruple pattern for the MOS), good (FLAG=0) events were used
with response matrices generated for each spectrum.  The spectra were binned
with a minimum of 20 counts per bin.  The final X-ray source positions were
determined after cross-correlation with the USNO A2.0 optical catalogue
based on the SAS task \emph{eposcorr} \citep[see][]{Tedds:2000}.

\begin{figure}
 \includegraphics[angle=-90,width=\columnwidth,clip=]{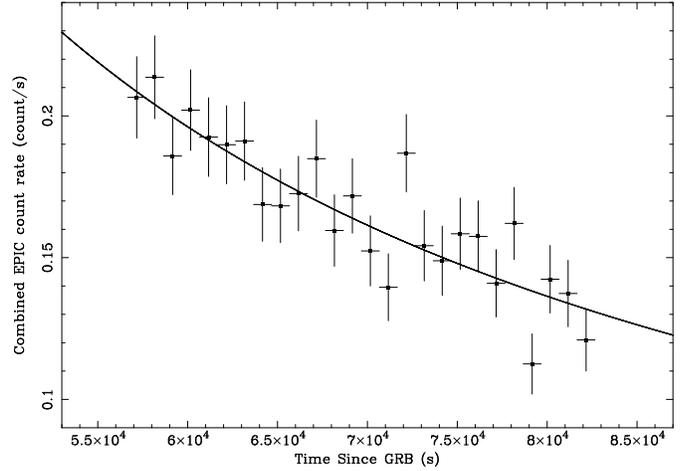}
 \caption{Combined EPIC-pn and MOS lightcurve of the 0.2--12\,keV
          afterglow of GRB\,020322.  The best-fit power-law decay ($\chi^2 =
          25.2$ for 24 degrees of freedom) is plotted and has a decay index
          of $1.26\pm0.23$.}
 \label{lightcurve}
\end{figure}

Three sources were detected in the \emph{BeppoSAX}-MECS error-circle
(Fig.~\ref{image}).  The brightest, with coordinates RA = 18h\,01m\,03.1s,
Dec. = +81\degr\,06\arcmin\,27\arcsec.9, and a 68\% error radius of
0.5\arcsec,\ (source `A' in Fig.~\ref{image}) was seen to fade
(Fig.~\ref{lightcurve}), with a probability of constancy of
$6\times10^{-12}$.  It is interesting to note the very good correspondence
($<0.5$\arcsec\ separation) between this source position and that of the
optical transient of \citet{2002GCN..1296....1B}, at RA = 18h01m\,02.98s
$\pm~0\arcsec.30$, Dec. = +81\degr\,06\arcmin\,28\arcsec.17
$\pm~0\arcsec.35$.

A power-law decay ($F\propto t^{-\beta}$) with index
$\beta=1.26\pm0.23$ fits the lightcurve well ($\chi^2_\nu=1.05$).  We
identify this source as the afterglow of GRB\,020322.  The Galactic
hydrogen absorbing column in this direction is $4.6\times10^{20}$\,cm$^{-2}$
\citep[using the FTOOL \emph{nh}]{1990ARA&A..28..215D}.  Its mean observed
0.2--10.0\,keV flux was $3.5\times10^{-13}$\,erg\,cm$^{-2}$\,s$^{-1}$,
making it the brightest GRB afterglow observed by \emph{XMM-Newton} so far,
roughly comparable in source counts to the afterglow of GRB\,011211
\citep{2002Natur.416..512R,astro-ph/0206480}. The other sources (`B' and `C'
in Fig.~\ref{image}) are both on the edge of the MECS error-circle, are
fainter (`B' = $6.3\pm0.5\times10^{-14}$\,erg\,cm$^{-2}$\,s$^{-1}$ and `C' =
$2.5^{+0.3}_{-1.0}\times10^{-14}$\,erg\,cm$^{-2}$\,s$^{-1}$) and show no
evidence for variability (null hypothesis probabilities of 0.95 and 0.69
respectively).

%The Optical Monitor did (not ??) detect an optical counterpart.  The
%sources were not detected with the Reflection Grating Spectrometers (RGS).

%
%--------RESULTS--------------
%
\section{Results\label{results}}
\begin{figure}
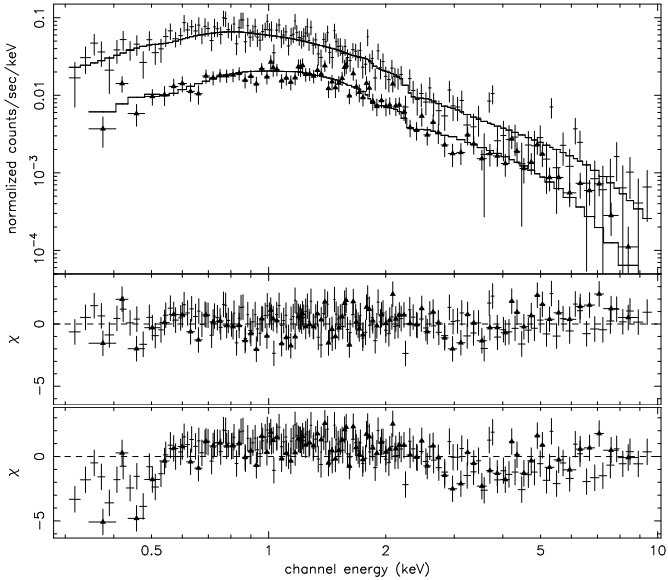

 \includegraphics[angle=-90,width=\columnwidth,clip=]{Ei272_f3a.ps}
 \includegraphics[angle=-90,width=\columnwidth,clip=]{Ei272_f3b.ps}
 \caption{EPIC-pn (crosses) and combined EPIC-MOS (triangles) spectrum of
          the afterglow of GRB\,020322 fit with a power-law absorbed by the
          Galactic column ($4.6\times10^{20}$\,cm$^{-2}$) and a redshifted
          neutral absorber. The residuals to this fit are shown in the
          middle panel. This model fits the data adequately
          ($\chi^2_\nu=1.04$).  Residuals to a model fit involving a
          power-law absorbed only by the Galactic column are plotted in the
          bottom panel.  This model does not fit the data well
          ($\chi^2_\nu=1.87$).}
 \label{spectrum}
\end{figure}

The complete EPIC spectrum (Fig.~\ref{spectrum}) is not well fit with a
Galactic-absorbed power-law model, $\chi^2 = 437.4$ for 234 degrees of
freedom (DoF).
Adding redshifted neutral absorption, the fit becomes acceptable
($\chi^2$/DoF = 241.5/232), equivalent to a best-fit local excess hydrogen
column density of $1.6_{-0.2}^{+0.3}\times10^{21}$\,cm$^{-2}$, however there
are no significantly detected absorption edges in the spectrum, indicating
that the redshift is probably greater than 0.7 (the 90\% confidence limit,
derived largely from the lack of detection of the neutral Oxygen absorption
edge), with a best-fit redshift of $1.8_{-1.1}^{+1.0}$.  The best-fit
neutral absorbing column density at $z=1.8$ is
$1.3\pm0.2\times10^{22}$\,cm$^{-2}$.  It is possible that the metal
abundance in the absorbing gas is low, which would also explain the lack of
absorption edges \citep[this seems unlikely however, given the probability
that GRBs occur in star-forming regions, e.g.][]{1999A&A...344L..67H}. At
$z=0$, abundances $<0.4$ times the solar values are required to fit the
data.
% The chi-squared space is non-monotonic.

It has been posited that GRBs in dense star-forming regions will photoionise
the circumburst medium \citep{1999A&A...343..111B}, giving rise to a NEI
plasma.  Ionised absorption cannot be ruled out in this case, with an upper
limit of 140\,erg\,cm\,s$^{-1}$ to the ionisation parameter of the absorber;
a fit where the excess absorption is ionised, with variable iron abundance,
gives as good a fit as the neutral absorption case ($\chi^2$/DoF =
239.4/230).  The redshift in the case of the ionised absorber is still
constrained, primarily by the Fe-L shell and \ion{O}{viii}-K absorption
edges. Fixing the ionisation parameter at the upper limit gives a 90\%
confidence interval for the redshift of 2.4--2.9.  In the case where the
absorber is not in equilibrium \cite[see for instance the marginal detection
of, and possible explanations for, a transient absorption feature
in][]{2000Sci...290..953A,2002MNRAS.330..383L}, the range of ionization
states will be increased, however it is the non-detection of an absorption
feature that is the primary lower constraint on the redshift, implying that
the redshift limits still hold in the case of a non-equilibrium absorber.

An absorption edge is marginally detected ($\chi^2$/DoF = 233.5/230,
$\gtrsim98$\% significance) at 2.8\,keV in the observed frame.  Attributing
this to the K-edge of neutral iron (at 7.1\,keV) implies a redshift of 1.6,
neutral Co (at 7.7\,keV) implies $z = 1.8$, neutral Ni (at 8.3\,keV) implies
$z = 2.0$; all of which are within the $1\,\sigma$ error bounds of the
redshift determination from the fit to the absorption.  Allowing the
abundance of iron to vary in the model excess absorber however, does not
improve the fit significantly.

% [z=1.0--2.2].
There is no evidence for any significant detection of emission features in
the spectrum.  The best-fit parameter values are presented in
Table~\ref{fit_pars}.

In order to assess spectral variability, the data were divided in two parts. 
The data were extracted from the first 11\,ks and from the remaining 15\,ks
in order to have roughly equal numbers of source counts in each spectrum.
The start time of the pn exposure was used as the start time for both the
MOS and pn datasets.  There is no significant difference between the
best-fit parameters for the first and second part spectra (Table~\ref{fit_pars}).
\begin{table}
 \caption{Best-fit parameter values fitting a Galactic-absorbed power-law
          with variable redshifted neutral absorption to a) the complete
          EPIC dataset, b) the first 11\,ks of exposure and c) the remaining
          15\,ks of exposure.  Ninety percent confidence intervals are
          quoted in parentheses under the relevant value.}
 \label{fit_pars}
 \begin{tabular}{@{}lcccc@{}}
\hline\hline
		& Obs. Flux	& $\Gamma$	& $z$		& N$_{\rm H}$ at $z=0$\\
\multicolumn{3}{c}{($10^{-13}$\,erg\,cm$^{-2}$\,s$^{-1}$)}& 	& ($10^{21}$\,cm$^{-2}$)\\
\hline
Complete	& 3.5		& 2.06		& 1.8		& 1.6\\
		& (3.3--3.7)	& (1.98--2.14)	& (0.7--2.8)	& (1.4--1.9)\\
0--11\,ks	& 3.8		& 2.1		& 0.20		& 1.7\\
		& (3.4--4.4)	& (2.0--2.3)	& (0.04--4.1)	& (1.3--2.1)\\
11--26\,ks	& 3.2		& 2.0		& 1.7		& 1.4\\
		& (3.0--3.5)	& (1.9--2.1)	& (0.6-- $>$$5$)& (1.1--1.8)\\
\hline
 \end{tabular}
\end{table}
The data were also divided into 5\,ks parts to assess spectral variability
on a shorter time scale; again there is no significant difference between
the spectra.

Given the recent results from \emph{XMM-Newton} implying that thermal
emission features may be common in early-time X-ray afterglows
\citep{2002A&A...393L...1W}, these data were tested to determine limits on
thermal emission with the plasma model used in \citet{2002Natur.416..512R}
and \citet{2002A&A...393L...1W}.  The addition of a collisionally-ionised
plasma model \citep[the `mekal'
model,][]{1985A&AS...62..197M,1995ApJ...438L.115L} to the absorbed power-law
model does not improve the fit significantly, ($\chi^2$/DoF = 239.1/229,
giving an $f$-test probability for the addition of three extra terms of
0.49) and yields significantly worse fits when fit instead of a power-law
($\chi^2$/DoF = 342.1/231).  Similar results are obtained in the
time-divided spectra.  The results of these fits are summarised in
Table~\ref{fit_chis}. Assuming a metal abundance nine times the solar value
and a plasma temperature of 4.1\,keV \citep[the best-fit values
from][]{astro-ph/0206480}, an upper-limit to any thermal emission of
$3.7\times10^{-14}$\,erg\,cm$^{-2}$\,s$^{-1}$ is determined, 11\% of the
total flux.

\begin{table*}
 \caption{Goodness of fit in terms of $\chi^2$ and numbers of degrees of
          freedom ($\chi^2$/DoF) to a) the complete EPIC dataset, b) the first
          11\,ks of exposure and c) the remaining 15\,ks of exposure, for six
          different models made up of three components: a power-law (PL),
          a redshifted neutral absorbing gas (Abs$_{\rm z}$), and a
          collisionally-ionised plasma (VMEKAL).  All models include
          absorption by the Galaxy ($4.6\times10^{20}$\,cm$^{-2}$).}
 \label{fit_chis}
 \begin{tabular}{@{}lcccccc@{}}
\hline\hline
Model		& PL	& Abs$_{\rm z}$+PL	& PL+VMEKAL	& Abs$_{\rm z}$+PL+VMEKAL	& VMEKAL	& Abs$_{\rm z}$+VMEKAL\\
\hline
Complete	& 437.4/234	& 241.5/232	& 364.4/230	& 239.1/229			& 398.6/232	& 342.1/231\\
0--11\,ks	& 203.5/116	& 123.9/114	& 155.3/112	& 117.5/111			& 179.0/114	& 162.5/113\\
11--26\,ks	& 189.0/123	& 109.4/121	& 151.6/119	& 108.4/118			& 161.6/121	& 138.6/120\\
\hline
 \end{tabular}
\end{table*}

%% Lightcurves -- variability and decay-rate.
%%
%% Spectra -- Best-fits, adding components.
%%
%% Temporal variability of spectra.
%%
%% Significance of results.

%
%--------DISCUSSION-----------
%
\section{Discussion\label{discussion}}

The afterglow of GRB\,020322 is the brightest observed to date by
\emph{XMM-Newton}, with $\sim30\%$ more source counts than the detection of
GRB\,011211.  It shows no evidence for line emission, permitting at most
$\sim10\%$ of the flux to come from a line-dominated thermal component similar
to that observed in GRB\,011211.  This result contrasts with the detection
of luminous emission lines and the good fit achieved with a thermal plasma
model to the three previous detections of GRB afterglows with
\emph{XMM-Newton} (GRB\,001025A, GRB\,010220 and GRB\,011211).  It is worth
noting that under similar time constraints (i.e.\ $\sim15$\,hours after the
burst) significant thermal emission is not detectable in the afterglow
spectrum of GRB\,011211 either.  It is possible therefore that thermal
emission could have dominated the early-time spectrum of the afterglow of
GRB\,020322 \emph{before} the time of this observation.  The uncertainty
surrounding the line-dominated component of the afterglow spectra (due to
the few available results) also allows us to speculate that the afterglow
spectrum of GRB\,020322 may have become line-dominated \emph{after} the end
of the \emph{XMM-Newton} observation.  The resolution of this question
requires continuous monitoring of GRB X-ray afterglows for as long as two
days after the burst.  This may soon be possible using a combination of
\emph{Swift} \citep{2000hgrb.symp..671G} and \emph{XMM-Newton}.

% or \emph{INTEGRAL} \citep{astro-ph/0205071}.

%INTEGRAL??

It seems very likely that GRBs are highly collimated, relativistic events,
with the beam angle constrained to less than $\sim30\degr$
\citep{2001ApJ...562L..55F,2001ApJ...554..667P,2002A&A...385..377Q}.  Bursts
observed with small viewing angles to the jet may have apparently brighter
synchrotron afterglows.  In this case, where the angle of the jet to the
line of sight is small and a brighter afterglow is observed, isotropic line
emission would be more difficult to detect.

A notable difference between this spectrum and that of GRB\,011211 is the
excess absorption required to fit the data in this case.  It is possible
that the absorbing gas is ionised.  However, in either case the redshift
must be high enough to shift the absorption edges out of the band (where the
gas has metal abundances similar to or greater than the solar value, a
reasonable assumption given that many GRBs are associated with strongly
star-forming regions, e.g. \citeauthor{1999A&A...344L..67H}
\citeyear{1999A&A...344L..67H}).

Extrapolating the best-fit unabsorbed X-ray (synchrotron) power-law model
into the optical R-band, gives an upper limit to the unreddened magnitude of
$19.4\pm0.6$ \citep[though this extrapolation may be further complicated by
the presence of inverse Compton emission in the X-ray spectrum, see for
example][]{2001ApJ...559..123H}. The observation of
\citet{2002GCN..1298....1G} occurred during the
\emph{XMM-Newton} observation where they detect an optical transient with an
R-band magnitude of $23.80\pm0.30$.  This implies an observed-frame
extinction of, at most, A$_{\rm R}=4.4\pm0.9$, a degree of extinction at
least concurrent with the fact that some GRBs are associated with regions of
vigorous star-formation \cite[see for example][]{1999A&A...344L..67H}.  This
value is consistent with the detected X-ray column density at $z=1.8$,
assuming a Galactic gas-to-dust ratio.

The detection of a potential host galaxy with a magnitude of $\sim27$
\citep{2002GCN..1536....1B} is also consistent with the redshift range
derived from the X-ray absorbing column, assuming a distribution of host
galaxy magnitudes similar to either the Hubble Deep Field North
\citep{2000ApJ...538...29C} or a sample of GRB host galaxies
\citep{2002AJ....123.1111B}.

%%(Does the 2.8keV edge-depth correspond to the absorbing column?)

%% Describe different model possibilities;
%% (1) Recent SN shell-heating
%% (2) Reflection/recombination models
%%     (instant SN)
%%
%% Do calculations for these models from the 
%%    spectral limits.
%%
%% What does this imply for the different
%%   models?

%
%--------CONCLUSIONS----------
%
\section{Conclusions\label{conclusions}}

\emph{XMM-Newton} detected the afterglow of GRB\,020322 with a mean observed
0.2--10.0\,keV flux of $3.5\pm0.2\times10^{-13}$\,erg\,cm$^{-2}$\,s$^{-1}$,
making it the brightest X-ray afterglow observed with this satellite so far. 
Fitting a power-law to the lightcurve yields a best-fit decay index of
$1.26\pm0.23$.  The spectrum -- the best quality spectrum of a GRB X-ray
afterglow to date -- is well fit with a power-law absorbed with neutral or
ionised gas significantly in excess of the Galactic column at redshift
$1.8_{-1.1}^{+1.0}$ and/or with low metallicity.  No emission features are
detected in the spectrum and a thermal (mekal) model fits the data poorly,
the upper limit on its contribution to the spectrum (assuming a 4\,keV
plasma with nine times the solar metal abundances) is
$3.7\times10^{-14}$\,erg\,cm$^{-2}$\,s$^{-1}$ ($\sim10\%$ of the total flux),
indicating that if a thermal component was present in the early-time
afterglow, it faded below a detectable level within the first
$\sim15$\,hours, making it much fainter than the synchrotron emission.

%% What data do we have?
%%
%% What have we seen in these data?
%%
%% Which is our preferred model to fit 
%%   these data?
%%
%% Can we rule out anything?

\begin{acknowledgements}
This work is based on observations obtained with \emph{XMM-Newton}, an ESA
science mission with instruments and contributions directly funded by ESA
Member States and the USA (NASA).
\end{acknowledgements}

\small
\bibliography{mnemonic,grbs}

\end{document}